\documentclass[prd,11pt, a4paper,
nofootinbib,
preprintnumbers,
]{revtex4}
\usepackage{epsfig}
\usepackage{graphicx}
\usepackage[normal]{subfigure}
\usepackage{rotating}

\begin{document}
\title{Non-unitarity effects in the minimal inverse seesaw model}
\date{September 10, 2009}
\author{Michal Malinsk\'{y}}\email{malinsky@kth.se} \thanks{\\ In collaboration with Tommy Ohlsson and Zhang He (KTH Stockholm) and Zhi-zhong Xing (IHEP Beijing). \\Talk given at {\it The 2009 Europhysics Conference on High Energy Physics,
		 July 16 - 22 2009
		 Krakow, Poland}}
\affiliation{Theoretical Particle Physics Group,
Department of Theoretical Physics,
Royal Technical Institute (KTH),
Roslagstullsbacken 21,
SE-106 91 Stockholm, Sweden.}
\begin{abstract}
A minimal version of the inverse seesaw model featuring only two pairs of  TeV-scale singlet neutrinos is discussed from the perspective of non-standard neutrino interactions. A particular attention is paid to the non-standard patterns of flavour and CP violation emerging due to the possibly enhanced non-decoupling effects of the heavy sector and the associated non-unitarity of the effective lepton mixing matrix.
\vspace*{0ex}
\end{abstract}
\maketitle
\maketitle
\section{Introduction}
With the upcoming generation of dedicated precision neutrino experiments and a continuing data-taking from the ongoing facilities the next decade offers very good prospects of measuring some of the observables underpinning the flavour violationg in the  lepton sector to an unprecedented accuracy level. Improving sufficiently our current knowledge about the neutrino oscillation parameters one can in principle even attempt to ask to what extent the current picture of three flavours oscillating into each other during the neutrino propagation from the source to the detector is complete or if there could be any substential room for physics beyond the standard framework. 
Indeed, the seesaw mechanism furnishing perhaps the most compelling understanding of the very tiny light neutrino mass scale relies on new dynamics coming to play at a certain scale, which could exhibit itself as various non-standard effects in the neutrino behavior.
Although the standard lore suggests such a new physics scale should be presumably very high (typically at around $10^{13}$ GeV complying nicely with the unification paradigm) there are theoretical scenarios that admit bringing it down into the TeV domain. This would not only provide a very attractive option to probe the underlying dynamics directly at the LHC but also give us a chance to observe various non-decoupling effects of this kind of a new physics relatively close to the electroweak scale.

\section{Non-unitary lepton mixing and neutrino oscillations}
One of the most profound implications of such scenarios is the possibility that the mixing matrix entering the leptonic charged-current interactions need not be unitary, i.e. the transformation matrix $V_{\alpha i}$ connecting the flavour ($\alpha$) and mass ($i$) bases $|\nu_{\alpha}\rangle=V_{\alpha i}|\nu_{i}\rangle$, when contracted only to the dynamically accessible light degrees of freedom, {\it does not} obey the unitarity relation $(VV^{\dagger})_{\alpha\beta}={\bf 1}$. In such case the traditional survival/transition probability formula receives the following form \cite{Antusch:2006vwa}:
\begin{equation}\label{eq:bigformula}
P_{\alpha\to \beta}\approx\sum_{i,j}{\cal F}_{\alpha\beta}^i{\cal F}_{\alpha\beta}^{j*}-4\sum_{i>j}{\rm Re}({\cal F}_{\alpha\beta}^i{\cal F}_{\alpha\beta}^{j*})\sin^2\left(\frac{\Delta m_{ij}^2 L}{4E}\right)+2\sum_{i>j}{\rm Im}({\cal F}_{\alpha\beta}^i{\cal F}_{\alpha\beta}^{j*})\sin\left(\frac{\Delta m_{ij}^2 L}{2E}\right)
\end{equation}
where ${\cal F}_{\alpha\beta}^{i}\equiv \sum_{\gamma\rho}(R^{*})_{\alpha\gamma}(R^{*})^{-1}_{\rho\beta}U_{\gamma i}^{*}U_{\rho i}$ with $R_{\alpha\beta}\equiv (1-\eta)_{\alpha\beta}/[(1-\eta)(1-\eta^{\dagger})]_{\alpha\alpha}$ and $\eta$ denoting the ``departure'' of $V$ from unitarity: $V=(1-\eta)U$. The first two terms in (\ref{eq:bigformula}) correspond to the so called zero-distance effects and the ``standard'' oscillation term respectively (giving rise to the traditional simple oscillation formula for a unitary $V$ in the $\eta\to 0$ limit) while the last term leads to extra CP violating effects for $\eta\neq 0$. The entries of the $\eta$ matrix are constrained from various experimental inputs like e.g. rare leptonic decays, invisible $Z$-boson width, neutrino oscillations etc. For illustration let us quote the 90\% C.L. limits given in \cite{Antusch:2008tz}: $|\eta_{ee}|\lesssim 2.0\times 10^{-3}$, $|\eta_{\mu\mu}| \lesssim 8.0\times 10^{-4}$, $|\eta_{\tau\tau}| \lesssim 2.7\times 10^{-3}$, $|\eta_{e\mu}| \lesssim 3.5\times 10^{-5}$, $|\eta_{e\tau}| \lesssim 8\times 10^{-3}$, $|\eta_{\mu\tau}| \lesssim 5.1\times 10^{-3}$.
\section{Non-unitarity effects in seesaw models}
Let us now inspect in brief the prospects of accommodating non-negligible non-unitarity effects within the three ``canonical'' realizations of the seesaw idea:\\
{\bf Type-I:}
In type-I seesaw, the structure $F\equiv M_{D}M_{R}^{-1}$ (with $M_{D}$ and $M_{R}$ denoting the Dirac and Majorana masses respectively) determining the overall scale of the $\eta$-matrix obeying 
$
\eta=\frac{1}{2}FF^{\dagger}
$
governs also the scale of the light neutrino masses $m_{\nu}=FM_{R}^{-1}F^{T}$.
Thus, non-unitarity effects can be sizeable only if $M_{D}$ is relatively close to $M_{R}$. This, however, implies a  need for  a severe structural fine-tuning in order to bring the eigenvalues of $m_{\nu}$ down to the observed level. A detailed discussion of this option can be found for instance in \cite{Kersten:2007vk}. However, if $M_{R}$ is at around $10^{13}$ GeV the heavy sector essentially decouples and the non-unitary effects are extremely tiny.\\
{\bf Type-II:} The ``canonical'' type-II realization of the seesaw mechanism does not bring in any extra sector that could mix with the SM leptons; the unitarity of the mixing matrix remains unaffected.\\
{\bf Type-III:}
The situation in the type-III seesaw is a close analogue of the type-I case with the only difference that $SU(2)_{L}$-triplet fermions are employed in type-III rather than singlets in type-I.
\subsection{Inverse seesaw}
Thus, in order to accommodate naturally the sub-eV light neutrino masses together with potentially sizeable non-unitary effects in the lepton sector one has to go beyond the simple ``canonical'' realizations of the seesaw idea. In particular, there is a need to disentangle the scale providing the suppression for the neutrino masses (like $M_{R}$ in the type-I \& III seesaw) from the scale entering $F$. 

To our best knowledge, the simplest model of this kind can be constructed along the lines of the inverse seesaw \cite{Mohapatra:1986bd} approach in which a third scale (usually denoted by $\mu$ besides $M_{D}$ and $M_{R}$) associated to an extra SM-singlet sector enters the full neutrino mass-matrix: 
\begin{equation}
M_{\nu}=\left(\begin{array}{ccc}
0 & M_{\rm D} & 0 \\
. & 0 & M^T_{\rm R} \\
. & . & \mu
\end{array}\right).
\end{equation}
where the undisplayed entries follow from symmetries of $M_{\nu}$.
One of the virtues of this strucure consists in the fact that $\mu$ is the only lepton-number-violating  mass term around and as such can be made arbitrarily small with no implications for  naturalness (in the t'Hooft sense, c.f. \cite{tHooft:1979bh}). 
For $\mu\ll M_{D}< M_{R}$  the neutrino spectrum consists of three light Majorana neutrinos with a mass matrix
$m_{\nu}\simeq M_{\rm D} M^{-1}_{\rm R} \mu (M^T_{\rm R})^{-1} M_{\rm
D}^{T}  = F \mu F^T$, and (usually) three pairs of heavy ($\sim M_{R}$) almost degenrate  Majorana neutrinos with oposite CP (and mass splitting proportional to $\mu$) which can be viewed as (three) heavy pseudo-Dirac neutrinos. It can be shown that the heavy states are admixed into the light flavours with an effective ``strength'' proportional to $F$ and the 3$\times$3 sub-matrix of the full (usually 9$\times$9) neutrino sector diagonalization matrix (in the basis of diagonal charged leptons) reads again $V \simeq
\left({\bf 1}-\frac{1}{2}FF^\dagger\right) U$, see e.g. \cite{Schechter:1981cv}, and the basic ``type-I'' formula $\eta=\frac{1}{2}FF^\dagger$ remains intact. Thus, while the non-unitarity effects are driven (as in the type-I case) by $F=M_{D}M_{R}^{-1}$, one has an option here to pull $M_{R}$ down to lower scales, perhaps even into the TeV region, because in such case the neutrinos can be kept naturally light if $\mu$ falls into about the keV domain.
 
The non-unitarity effects emerging in the original inverse seesaw with three pseudo-Dirac heavy neutrinos have been discussed at length in a recent paper \cite{Malinsky:2009gw} so here we will confine ourselves only to the salient points of the analysis.
In full generality, one can parametrize the $F$ matrix (\`{a} la Cassas-Ibarra \cite{Casas:2001sr}) in terms of the measurable light neutrino masses grouped into a diagonal matrix $d_{\nu}$, a complex orthogonal matrix $O$ and the observed lepton mixing matrix $U$ (including possible Majorana phases) so that  
$
\label{eq:eta}
\eta=\frac{1}{2}FF^\dagger =\frac{1}{2}U \sqrt{d_{\nu}} O  \mu^{-1} O^\dagger
\sqrt{d_{\nu}} U^\dagger \ .
$
Interesting non-trivial correlations between the entries of $\eta$ emerge in special cases in which further assumptions are made about the shape of $O$ and/or $\mu$ matrices (reflecting various physically interesting situations at the level of an underlying theory). For example, for a flavour-blind $\mu=\mu_{0}\times {\bf 1}$ and arbitrary $O$, the relevant formula reads
$\eta = \frac{1}{2} \mu_0^{-1} U \sqrt{d_\nu} {\rm exp}
(2\,{\rm i}A) \sqrt{d_\nu} U^\dagger$ with $A$ being a real
antisymmetric matrix \cite{Pascoli:2003rq} parametrized by three real numbers only. This degeneracy admits to cast more stringent bounds  on $|\eta_{e\tau}| < 2.3 \times
10^{-3}$ and $|\eta_{\mu\tau}| < 1.5 \times 10^{-3}$ while saturating the 90\% C.L. limits for the others. 
\subsection{Minimal inverse seesaw \& non-unitarity effects}
The situation becomes even more interesting in the minimal inverse seesaw (MISS) framework \cite{Malinsky:2009df} with only two pairs of heavy pseudo-Dirac neutrinos around the TeV scale. This, indeed, is enough to generate two non-zero light neutrino mass-squared-differences while one of the light neutrinos remains massless.
Due to the $2\times 2$ shape of the $\mu$ and $M_{R}$ matrices in this case $\eta$ is obtained by replacing the original ($3\times 3$ complex orthogonal) matrix $O$ above by a rectangular ($3\times 2$ complex) matrix $R$ instead: $\eta=\frac{1}{2}U \sqrt{d_{\nu}} R  \mu^{-1} R^\dagger
\sqrt{d_{\nu}} U^\dagger $, which can be parametrized by means of a single (complex) angle\footnote{The shape of $R$, however, differs for normal and inverted hierarchy of the light neutrino spectrum; an interested reader is deferred to the original work \cite{Malinsky:2009gw} adopting the parametrization from \cite{Ibarra:2003up} for further details.}.  Expanding in powers of $\varepsilon \equiv({\Delta
m^2_{21}/|\Delta m^2_{31}|})^{1/4} \simeq 0.42$ one obtains the following approximate formula for the $\eta$ matrix:
\begin{equation}\label{eq:etaMISS}
\eta\propto\left(\begin{array}{ccc}
{\cal O}({|x|}) & - x \cos\theta_{23}& x\sin\theta_{23} \\
.&  \cos^{2}\theta_{23} & \pm  \frac{1}{2}\sin 2\theta_{23} \\
. & . & \sin^{2}\theta_{23}
\end{array}\right)+{\cal O}(\varepsilon^{2-x}) .
\end{equation}
Here $x\in\{1,0\}$ where the former value (the upper sign above) corresponds to the normal hierarchy of the neutrino spectrum while the latter (the lower sign in (\ref{eq:etaMISS})) to the inverted hierarchy case.

\begin{figure}[t]
\begin{center}
\includegraphics[width=8cm]{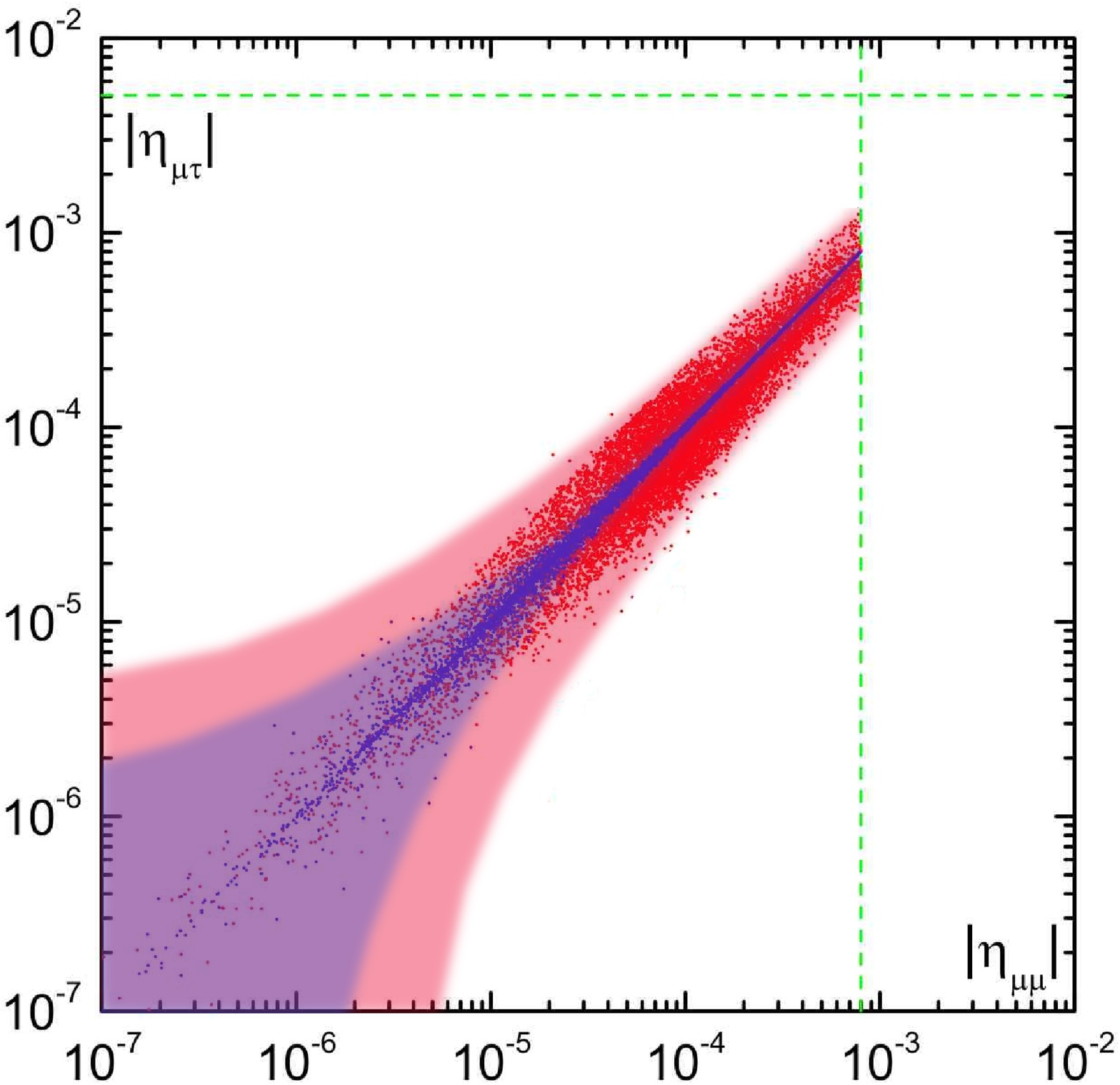}
\includegraphics[width=8cm]{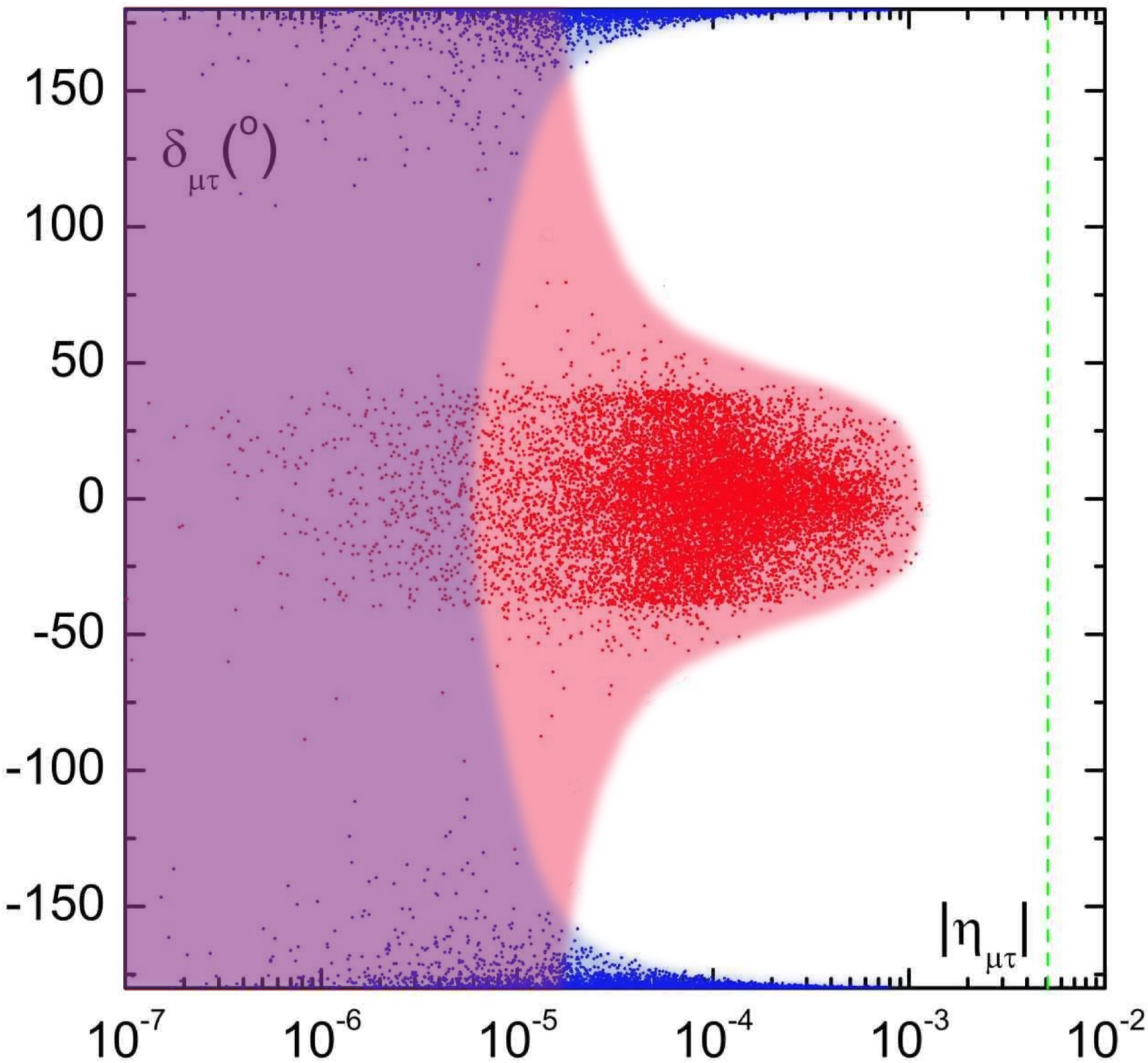}
\caption{\label{fig:etamutau} Correlations among various parameters
governing the non-unitarity effects in the MISS. Red points (simulated) and regions (extrapolated) correspond to
the normal neutrino mass hierarchy, while the blue ones stand for
the inverted mass hierarchy. Generic experimental
constraints are indicated by the green dashed lines. For sake of 
simplicity, the correlations in their determination have
been neglected).} 
\end{center}
\end{figure}

The structure (\ref{eq:etaMISS}) has two immediate implications: 1) Due to the observed approximate maximality of the atmospheric mixing one has $|\eta_{\mu\mu}|\simeq |\eta_{\mu\tau}|\simeq|\eta_{\tau\tau}|$ and
$|\eta_{e\mu}|\simeq |\eta_{e\tau}|$ which is a genuine testable prediction of the MISS model. The correlation is stronger in the inverse hierarchy case because the sub-leading term in (\ref{eq:etaMISS}) is smaller. 
2) Only the $|\eta_{\mu\tau}|$ off-diagonal entry can be sizeable, which yields a rather specific pattern of non-standard flavour and CP violation effects potentially emmerging in the current scenario. Moreover, in the inverse hierarchy case the phase of  $\eta_{\mu\tau}$ is nearly maximal (corresponding to the minus sign of the 23 entry of the leading term in (\ref{eq:etaMISS}) perturbed only at the ${\cal O}(\varepsilon^{2})$ level while for the normal hierarchy it is preferably small (zero at the leading level perturbed by the subleading ${\cal O}(\varepsilon)$ terms).
One can observe this kind of behavior in Figure \ref{fig:etamutau} depicting the results of the numerical analysis\footnote{Let us remark that for very small values of $|\eta_{\alpha\beta}|$ the uncertainities in determination of the neutrino mass-square-differences and the lepton mixing angles dominate over the effects subsumed by formula  (\ref{eq:etaMISS}) which leads to a typical smearing of the patterns obtained from a full numerical simulation towards the $|\eta_{\alpha\beta}|\to 0$ limit, c.f. Figure \ref{fig:etamutau}.}. 
Concerning the prospects of measuring the value of $|\eta_{\mu\tau}|$ at the near future experimental facilities, one can conclude that an ideal tool for that would be a neutrino factory with an OPERA-like near detector  within up to few tens of kilometers from the source. In such case an aparatus with 5kt fiducial volume exposed by neutrinos from 5+5 years of  muon+antimuon runs (corresponding to  around $10^{21}$ useful muon decays) could probe $|\eta_{\mu\tau}|$ down to the level of about 5$\times 10^{-4}$. There are also reasonable prospects of observing extra CP features in such situation provided the phase of $\eta_{\mu\tau}$ is around $\delta_{\mu\tau}=\pm\pi/4$ (otherwise the third (CP) term in formula (\ref{eq:bigformula}) is further suppressed by a small product $|\eta_{\mu\tau}| \sin\delta_{\mu\tau}$). For more details an interested reader is deferred to the original work  \cite{Malinsky:2009gw}.

\section{Conclusions}
The 'traditional' seesaw models of neutrino masses often resort on a directly inaccessible dynamics in the vicinity of the unification scale. However, in many extended scenarios the new physics scale can be as low as few TeV and the enhanced non-decoupling effects can be testable in the upcoming experimental facilities. From this perspective, the minimal inverse seesaw provides an appealing low-energy framework accommodating naturally the observed neutrino masses and mixings. Moreover, the distinct pattern of non-standard flavour and CP effects due to the non-unitarity of the lepton mixing matrix can be observable in the near future neutrino experiments.

\end{document}